\documentstyle[twoside,fleqn,espcrc2,epsf]{article}

\newcommand{\postscript}[2]
{\setlength{\epsfxsize}{#2\hsize}
\centerline{\epsfbox{#1}}}

\title{The neutrino mass and other  possible signals 
of lepton-number violation
in supersymmetric theories
\thanks{Based on a talk given at
SUSY '96, College Park, MD, May 29 - June 1, 1996.}
}

\author{Nir Polonsky
\address{Sektion Physik der Universit$\ddot{\mbox{a}}$t  
M$\ddot{\mbox{u}}$nchen, 
Theoretische Physik - Lehrstuhl Prof.\ Wess,\\
Theresienstrasse 37, D-80333 M$\ddot{\mbox{u}}$nchen,  Germany}
\thanks{Work supported by a fellowship of the DFG.  Address
after September 1, 1996:  Dept. of Physics and Astronomy, 
Rutgers University,
Piscataway, NJ 08855-0849, USA.}
}

\begin{document}

\begin{abstract}
We review a recently proposed framework in which
the neutrino mass is a signal of supersymmetry breaking
and is suppressed dynamically.
In addition, we briefly comment on some possible consequences
of general lepton-number violation in supersymmetric theories, 
{\it e.g.,} dijet and multijet signals and $jj \rightarrow ll\gamma\gamma$. 
\newline \newline
LMU-TPW-96-16

\end{abstract}

\maketitle

{\bf 1.} Lepton number violation in supersymmetric models
is only mildly constrained (see, $e.g.$, Bhattacharyya's contribution),
allowing for the generation
of a tree-level neutrino mass at the weak scale (see
Section 2),
and leading to non trivial signatures of supersymmetry
(see Section 3).
We discuss 
a few examples, stressing 
model-building and phenomenological aspects.

{\bf 2.} It has been pointed out by Hall and Suzuki \cite{hs}
that by explicitly introducing a $\Delta L = 1$ (where $L$ is
the lepton number) mass term in the superpotential, $i.e.$,
\begin{equation}
W = \mu_{H}H_{1}H_{2} + \mu_{L}LH_{2} + {\mbox{ Yukawa terms}},
\label{W1}
\end{equation}
one mixes the neutrinos and neutralinos, leading to a tree-level
mass for one neutrino species. 
(Here $L$ is a lepton doublet 
and $H_{1,2}$ are the hypercharge $Y = \mp 1$ Higgs doublets.)
The Higgs-lepton mixing is sufficient
 to generate an expectation value 
for the scalar neutrino (sneutrino) 
once electroweak symmetry is broken,
leading to additional gaugino-neutrino mixing. 
The two sources for the mixing, the $\mu$ 
parameter and the expectation
value, are, in fact, four-vectors in field space. 
They  explicitly and
spontaneously break the $SU(4)$ symmetry of 
$[H_{1},L_{\tau,\,\mu,\,e}]$
rotations (in field space) down to a residual  $SU(2)$.
Thus, two (neutrino) states remain massless at tree-level.
(However, Yukawa and Yukawa-gauge loops
explicitly break the residual $SU(2)$, and all states are massive
at the loop level.)

Here, we outline  a realization of that idea \cite{lmu9}  
within a framework
of a spontaneously broken $U(1)$ $R$-symmetry, and, in which
the neutrino mass is suppressed dynamically.
A detailed discussion,  additional examples, and a comprehensive
reference list, can be found in Ref.\ \cite{lmu9}.

{\bf 2.1.} {\it The $U(1)_{R}$ framework:}
A spontaneously broken $U(1)_{R}$ is often present in
models of dynamical supersymmetry breaking, and thus, it is
a natural candidate to set the selection rules for nonrenormalizable
operators in the low-energy superpotential. 
We will further assume that supersymmetry, 
as well as the $U(1)_{R}$, are broken in the hidden sector
($e.g.$, in gaugino condensation models) at a scale 
$\Lambda= {\cal{O}}(10^{11})$ GeV. 
The $R$-axion in this case could be, 
for example, invisible, or  heavy due to
a possible anomaly with respect to the hidden QCD group \cite{kim}.

Operators in the effective theory are suppressed by inverse powers
of the Planck mass, $M_{P}$. The possible ``non-singlet'' operators 
are the $\mu$ parameters with $R(\mu_{H}) = 2$ and  $R(\mu_{L}) = 1$,
and the lepton and baryon number violating Yukawa couplings with 
$R(h^{\mbox{\tiny LNV}},\, h^{\mbox{\tiny BNV}}) = -1$. 
(We define $R = 3B + L$ for the chiral superfields, 
and adopt  the normalization
$R(W) = 2$. $B$ and $L$ 
are  baryon and lepton number, respectively, and
$R = L$ for non-baryonic chiral superfields.)
Dimensional arguments suggest that the latter  are given by, $e.g.$,
$h^{\mbox{\tiny LNV}} \sim |\mu_{L}/M_{P}| \rightarrow 0$. 
For example, consider  a hidden superfield 
$Z = ( z, \tilde{z}, F_{Z})$ and $R(Z) = 1$.  
Then, $\mu_{H} = \langle z\rangle^{2}/M_{P}$ 
and   $\mu_{L} = F^{*}_{Z}/M_{P}$ 
are of the same order of magnitude $\sim \Lambda^{2}/M_{P}$, 
and the holomorphicity of the superpotential
implies that $h^{\mbox{\tiny L(B)NV}}$ 
are suppressed, as promoted above. 

The models\footnote{
The superpotential (\ref{W1}) contains only
$L$ and $B$ conserving ``Yukawa terms'' = $h_{U}H_{2}QU + 
h_{D}H_{1}QD + h_{E}H_{1}LE$ 
(where $U$, $D$, and $E$ are the quark
and lepton singlets, respectively, $Q$ are the quark doublets,
and we suppress family indices).} 
contain a very restricted form of lepton number violation,
$i.e.$, only in the supersymmetric mass terms.
Since $U(1)_{R}$ and the field-rotation $SU(4)$ do not
commute, it cannot be simply rotated (at high energies)
from the mass to the Yukawa operators.
Thus, there is a distinction between the high
and low-energy lepton number definitions. The former is defined
by the superpotential ($e.g.$, by the term $h_{D}H_{1}QD$).
The latter is defined, after weak-scale
rotations, $e.g.$, by requiring $\langle \hat{L_{i}}\rangle = 0$
(the caret denotes rotated fields).  
Note  that once (low-energy)
rotations are performed 
$(i)$ baryon number violation is still absent and
$(ii)$ the lepton-number violating Yukawa couplings 
are proportional to the ordinary Yukawa couplings.
Therefore, the lepton number conserving and 
violating Yukawa couplings are diagonalized simultaneously,
suppressing new contributions to FCNC's. 
Furthermore, 
the $h^{\mbox{\tiny LNV}}$  are naturally small
(and, in general, satisfy all constraints).
The superpotential (\ref{W1}) and 
the corresponding $U(1)_{R}$ framework
outlined above are quite striking. The only new supersymmetric mass 
is that of the neutrino superfield, and it is generated
naturally. In general, supersymmetric
mass parameters (which are not ${\cal{O}}(M_{P})$) 
represent a small perturbation at high energy 
and parameterize the high-energy physics.
The neutrino mass, which results here
from such a parameter,
is also a signal of that physics and, in particular, 
of supersymmetry breaking.

{\bf 2.2} {\it Dynamical suppression on the neutrino mass:}
The neutrino mass is constrained
(from energy density)
$m_{\nu} \leq {\cal{O}}(100)$ eV $\sim 10^{-9}M_{Z}$
(while  we expect $\mu =  {\cal{O}}(M_{Z})$).
Yet, there is no need to encode the  
${\cal{O}}(10^{-9})$ suppression factor 
in the superpotential, $i.e.$, 
in the ratio $\mu_{L}/\mu_{H}$.
{\bf The tree-level neutrino mass 
vanishes if the $\mu$ and expectation value
four-vectors are aligned in field space} (see also Ref.\ \cite{wis}).
In this case, the $SU(4)$ is broken down to only $SU(3)$, leaving
all three  neutrinos massless.
The alignment also enables one to consistently 
define the Standard-Model (SM) lepton number
(since $\mu_{L}$ and $\langle L \rangle$ would be rotated away 
simultaneously).

The merit of the above  mechanism is that
one need not impose the alignment. It can
be achieved dynamically. Let us assume
$(a)$ universal soft scalar masses $m_{0}^{2}\phi_{i}\phi_{j}^{*}\delta_{ij}$
at the grand scale 
(in fact, we need to require universality only
of lepton and Higgs masses\footnote{Note that this condition
is difficult to satisfy in unified models \cite{lmu9}.})
and $(b)$ the proportionality of the soft scalar mass terms
($m^{2}\phi_{i}\phi_{j} + h.c.$) 
to the respective $\mu$ parameters\footnote{A similar mechanism 
could operate in low-energy supersymmetry breaking models 
(if $B$-proportionality holds -- see Pomarol's contribution).}
($i.e.$, ``$B$ proportionality''). 
It is a straightforward exercise to renormalize
the model and solve the minimization equations.
However, the resulting alignment can also be understood
intuitively. Given our boundary conditions,
the $\mu$-vector is the only direction in field space 
up to perturbations proportional to $h_{D}^{2}/8\pi^{2}$
(from scaling). Thus, the expectation value must align
in this direction. 
The small perturbations, however,  create a tiny
misalignment that permits a small neutrino mass.

The alignment ($i.e.$, 
the (output) expectation value ratio $\langle L \rangle /
\langle H_{1} \rangle \equiv \nu_{L}/\nu_{H}$
$vs.$ the (input) $\mu_{L}/\mu_{H}$ ratio) 
in the models is shown in Fig.\ 1 
for $\tan\beta \equiv \langle H_{2} \rangle/ \langle \hat{H_{1}} \rangle
= 5, 15, 30$.  
(For simplicity, we considered the one generation case.) 
It is excellent for small $\tan\beta$ and 
can have different values for large $\tan\beta$.
(The dependence of $m_{\nu}$
on $\tan\beta$  is not trivial and is explored in \cite{lmu9}).
\begin{figure}[htb]
\postscript{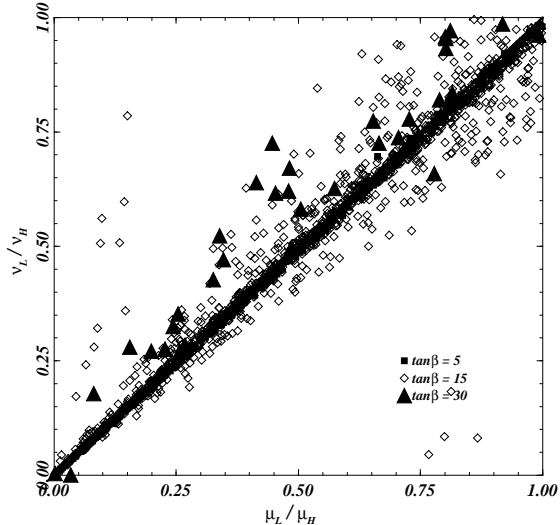}{1.0}
\caption{The dynamical alignment mechanism}
\label{fig:fig1}
\end{figure}
We find that, typically, 10 eV $\leq m_{\nu} \leq 10$ MeV.
The neutrino mass is sufficiently suppressed
due to the smallness of the  radiative corrections (to the universal
boundary conditions). The corrections only slightly perturb
the dynamical alignment, 
lifting the degeneracy by a small amount, and 
allowing for a neutrino mass of the correct magnitude.

{\bf 3.} In the remainder  of this contribution we would
like to comment on some possible signatures
of supersymmetric models with (low-energy) 
$h^{\mbox{\tiny LNV}}\neq 0$
($i.e.$, models with the $LQD$ and/or  $LLE$ operators).
(Section 2 represents a special case.)   
Typical signals, $e.g.$, like-sign dileptons,
are well studied (see, $e.g.$, V. Barger 
in these proceedings).
Here, we will comment  
on two (un-typical) examples:
The lightest supersymmetric
particle (LSP) is the photino and it decays radiatively,
or it is the sneutrino and it decays to $2j$.
(Note that quite generally the LSP is not a dark
matter candidate once lepton number is violated.) 
Our purpose is to stress  the wealth of new 
phenomena once lepton number is not conserved,
and to motivate further investigation.

{\bf 3.1} {\it Radiative decay of the photino:}
The photino LSP decays $\tilde{\gamma} \rightarrow
f\tilde{f} \rightarrow {2j+l}$ (where the tilde denotes superpartner, $j$
is a hadronic jet,
and we assumed dominance of the $LQD$ operator)
at tree level and  $\tilde{\gamma} \rightarrow \gamma\nu$ radiatively.
The former is expected to dominate unless
it is forbidden kinematically 
($i.e.$, $M_{\tilde{\gamma}} < 2m_{b}$
if the $LQ_{3}D_{3}$ operator is dominant)
or the photino is extremely light.
The tree level decay is, however, 
absent if the 
source of lepton number violation
is mainly the sneutrino expectation value \cite{daw}.
There could also be a situation 
in which the leading tree-level signal is hidden 
in the SM backgrounds while the radiative decay
produces a spectacular signature.
Of particular interest is the case
$jj \rightarrow \tilde{l}^{+}\tilde{l}^{-} 
\rightarrow l^{+}l^{-}2\gamma2\nu$.
(The tree level multijet signal is $l^{+}l^{-}2l4j$,
where $2l$ is a like-sign dilepton or neutrino $\rlap/E_{T}$.)
Such a scenario
may be able to provide a valid interpretation
to the $e^{+}e^{-}2\gamma +\rlap/E_{T}$ event observed by CDF
(see contributions by Thomas and by Kane).

{\bf 3.2} {\it The snuetrino decay to $2j$:}
If the sneutrino is the LSP, it would be singly produced by
either leptons or jets, and decay 
(assuming dominance of the $LQD$ operator) to $2j$.
The general enhancement of the dijet signal 
is discussed in \cite{lmu9}. 
It is ${\cal{O}}(10\%)$ of the QCD signal 
(at the threshold) \cite{dim}.
However, most probably the charged $SU(2)$ partner of the sneutrino
(and possibly  other scalars)
would decay similarly, leading to 
a larger (and smeared) enhancement 
of the dijet signal. 
Cascade decays lead, in this scenario,  
to multijet signals.

{\bf 4.} In conclusion, the supersymmetric extension
may not conserve lepton number.
In particular, if the neutrino mass
is (naturally) generated at low-energies
($e.g.$, from supersymmetric mass parameters).
The wealth of possible signals of supersymmetry,
in this case, requires further investigation.

{\bf  Acknowledgments:}
I am grateful to Hans-Peter  Nilles 
for his  collaboration,
and to Herbi Dreiner for the discussion of the photino decay modes.

\end{document}